\documentclass[aps,twocolumn,showpacs,superscriptaddress,citeautoscript,prb,floatfix, 10pt]{revtex4-2}

\usepackage{graphicx}
\usepackage{dcolumn}
\usepackage{bm}
\usepackage{color}

\usepackage{hyperref}
\usepackage{amssymb}
\usepackage{siunitx}
\usepackage{physics}
\usepackage{placeins}
\definecolor{brown}{rgb}{0.59, 0.29, 0.0}

\newcommand{\fig}{Fig.~{}}
\newcommand{\figs}{Figs.~{}}
\newcommand{\eqn}{Eq.~{}}
\newcommand{\eqns}{Eqs.~{}}

\newcommand{\jhat}{\boldsymbol{\hat{\jmath}}}

\newcommand{\hdirac}{\hat{H}_{\text{D}}}
\newcommand{\hblg}{\hat{H}_{\text{BLG}}}

\makeatletter
\newcommand\footnoteref[1]{\protected@xdef\@thefnmark{\ref{#1}}\@footnotemark}
\makeatother

\begin{document}

\title{Coherent suppression and dephasing-induced reentrance \\ of high harmonics 
	in gapped Dirac materials}

\author{Wolfgang Hogger}
 \email{wolfgang.hogger@ur.de}
\affiliation{Institute of Theoretical Physics, University of Regensburg, 93040 Regensburg, Germany}

\author{Alexander Riedel}

\affiliation{Institute of Theoretical Physics, University of Regensburg, 93040 Regensburg, Germany}

\author{Debadrito Roy}

 \affiliation{Indian Institute of Science, Bengaluru 560012, India}

\author{Angelika~Knothe}

\affiliation{Institute of Theoretical Physics, University of Regensburg, 93040 Regensburg, Germany}

\author{Cosimo Gorini}

\affiliation{
 SPEC,CEA, CNRS, Université Paris-Saclay, 91191 Gif-sur-Yvette, France
}

\author{Juan-Diego Urbina}

\affiliation{Institute of Theoretical Physics, University of Regensburg, 93040 Regensburg, Germany}

\author{Klaus Richter}

\affiliation{Institute of Theoretical Physics, University of Regensburg, 93040 Regensburg, Germany}

\date{\today}

\begin{abstract}
High-harmonic generation in solids by intense laser pulses provides a fascinating platform for  studying material properties and ultra-fast electron dynamics, where its coherent character is a central aspect. Using the semiconductor Bloch equations, we uncover a mechanism suppressing the high harmonic spectrum arising from the coherent superposition of intra- vs. inter-band contributions.
We provide evidence for the generality of this phenomenon by extensive numerical simulations exploring the parameter space
in gapped systems with both linear dispersion, such as for massive Dirac Fermions, and with quadratic dispersion, as e.g.\ for bilayer graphene.
Moreover, we demonstrate that, upon increasing dephasing, destructive interference between intra- and inter-band contributions is lifted. This leads to reentrant behavior of suppressed
high harmonics, i.e.\ a crossover from the characteristic spectral "shoulder" to a slowly decaying signal involving  much higher harmonics.
We supplement our numerical observations with analytical results for the one-dimensional case.
\end{abstract}

\maketitle

\section{Introduction}

High-harmonic generation (HHG) from solids has attracted considerable attention in recent years due to its potential to probe and manipulate electron dynamics on ultrafast timescales and with sub-wavelength spatial resolution \cite{Schubert2014,Hohenleutner2015}, as well as a promising platform for compact light sources in the ultraviolet or soft X-ray wavelength-regime \cite{T.T.Luu2015, M.Sivis2017, Seres2019}. The generation of high harmonics in solids is driven by the strong interaction of intense laser pulses with the material's electronic structure, leading to the emission of photons with energies corresponding to multiples of the driving laser frequency. The first experimental realization of HHG from solids in 2011 \cite{Ghimire2011} paved the way for understanding and controlling HHG in various materials such as wide-gap dielectrics \cite{Schubert2014, Hohenleutner2015, Du2018}, unstrained \cite{Yoshikawa2017, Sato2021, chizhova2017} and strained graphene \cite{Rana2024}, twisted bilayer graphene \cite{Du2021, Mrudul2024, Molinero2024}, topological insulators \cite{Schmid2021,Bai2020High-harmonicStates,Heide2022ProbingGeneration,Baykusheva2021}, strained TMDs \cite{Guan2019}, monolayer WS$_2$ \cite{kim2025} and semi-Dirac and Weyl materials \cite{Islam2018,Bharti2022High-harmonicSemimetal,Medic2024}.

\begin{figure}[htb]
\includegraphics[width=240pt]{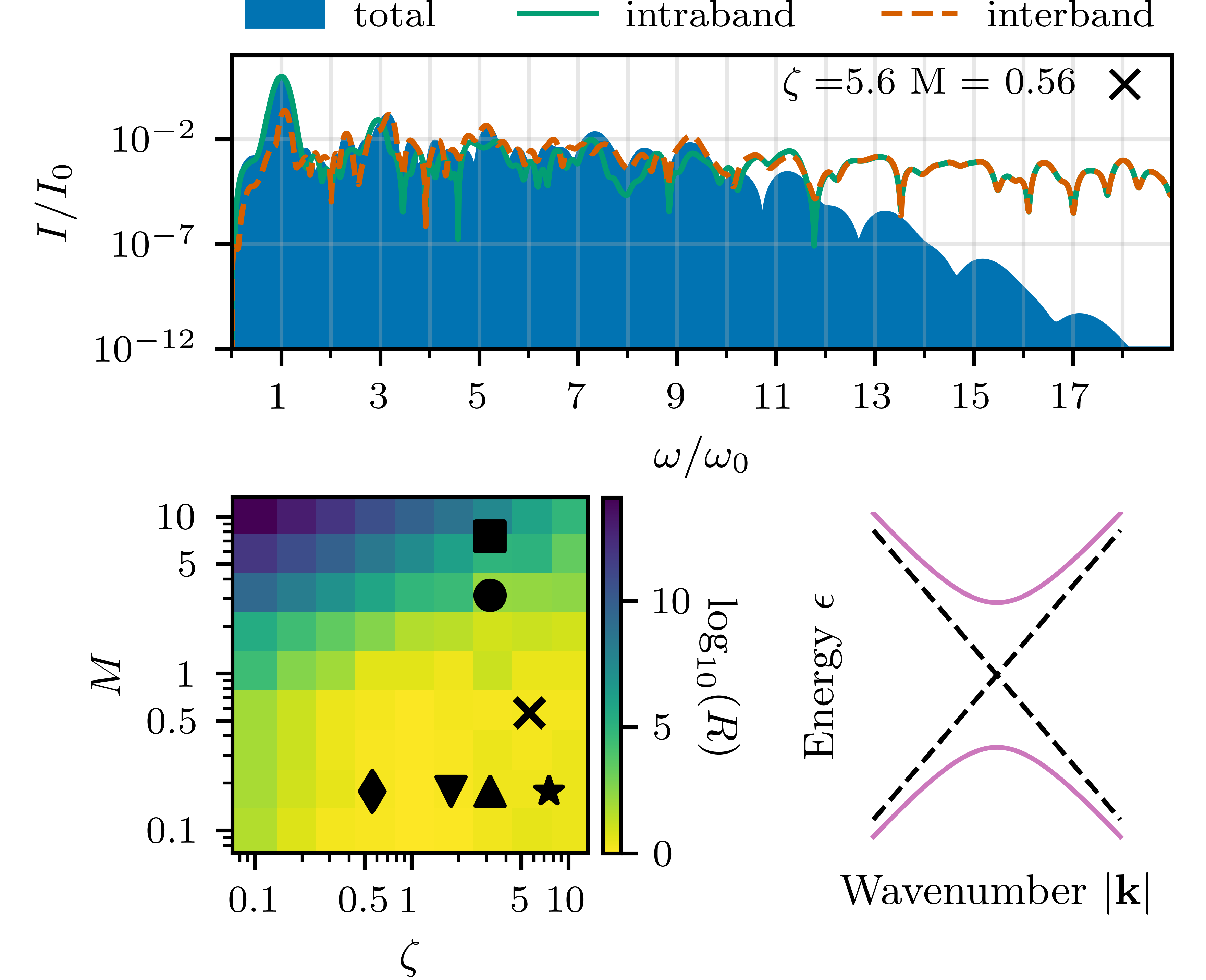}
\caption{\label{fig:fig1} Top: Destructive interference between the inter- and intra-band contributions to the HHG emission causes the total signal to be drastically reduced. Bottom left: Quantifying the degree of reduction of the total HHG signal by $R=\ev{I^{\text{inter}}/I^{\text{intra}}}_{\omega}$, we observe that coherent suppression is most efficient for small
 multi-photon numbers, $M$, and large strong-field parameters $\zeta$, i.e., in the regime of small gaps and strong driving fields (parameters defined in Eq.~(\ref{eq:parameters})). Markers refer to example spectra in the top panel and in \fig \ref{fig:fig2}. Bottom right: Schematic illustration of gapped Dirac band structure $\pm\sqrt{v_F^2k^2+\Delta^2}$ (solid purple lines) and diabatic energies $\pm v_F k_x$ (dashed).}
\end{figure}

In atomic gases,  the three-step recollision model \cite{Krause1992,Corkum1994,Lewenstein1994} provides a well-established theoretical framework to understand the underlying mechanisms of HHG. Solid-state HHG is more subtle due to the periodic crystal lattice and (multiple) electronic bands.
Its microscopic mechanism can be understood as the interplay between coherent inter-band polarization and intra-band dynamics that
is central to this work (see \fig \ref{fig:fig1}).
Both processes originate from the injection of a valence electron into an unoccupied state in the conduction band during a fraction of an optical cycle.
The intra-band contribution arises from population dynamics and does not depend on phase coherence between bands, whereas the inter-band current depends explicitly on electron-hole coherence.

Ghimire et al. \cite{Ghimire2011} suggested the intra-band current to be the primary source of HHG, whereas Schubert et al. \cite{Schubert2014} consider the combined action of dynamical intra-band Bloch oscillations and coherent inter-band excitations as the physical origin.
Studies with graphene have shown that intra-band \cite{Mrudul2021Light-inducedGraphene} or inter-band \cite{Mrudul2021High-harmonicGraphene} can dominate under certain conditions, strongly depending on laser parameters.
The three-step model was adapted to solid-state HHG in \cite{Vampa2014,Vampa2015}, which has since been applied to various scenarios with different modifications \cite{Zurron2018TheoryGraphene,Kruchinin2018,Yue2020,Yue2021,Parks2020WannierSemiconductors}.
However, it relies on various assumptions like low valence band depletion and inter-band dominance in its current form.

\FloatBarrier
Here we systematically elaborate on the interplay of intra- and inter-band dynamics in HHG.
We show that and explain why, in relevant parameter regimes, intra- and inter-band contributions cancel coherently due to destructive interference, leading to a suppressed HHG signal, cf.~\fig\ref{fig:fig1}.
We demonstrate the generality of our findings by comparing two model systems: massive Dirac fermions,  a prototypical model for topologically non-trivial matter, and a model with quadratic dispersion.
Similar results were numerically observed in gapped graphene \cite{Murakami2022},

and  \cite{Wilhelm2021} suggested this cancellation as a hallmark of linear dispersions.
We present extensive numerical HHG data

based on the Semiconductor Bloch Equations (SBE) and provide a microscopic understanding using analytical perturbation theory.
The latter presents a new approach in a regime where the three-step model is not applicable.

We further demonstrate that dephasing counteracts the suppression effect. Notably, we find a re-entrant {\it increased} HHG signal at large frequencies for {\it decreasing} dephasing times $T_2$ and show that the high harmonics intensity results from a power law
$\sim(\omega T_2)^{-1}$ in the relative inter- and intra-band phase.
This manifests as a characteristic dependence of re-entrant harmonics on dephasing strength, which may provide an experimentally accessible signature.

\begin{figure*}[t]
    \centering
    \includegraphics{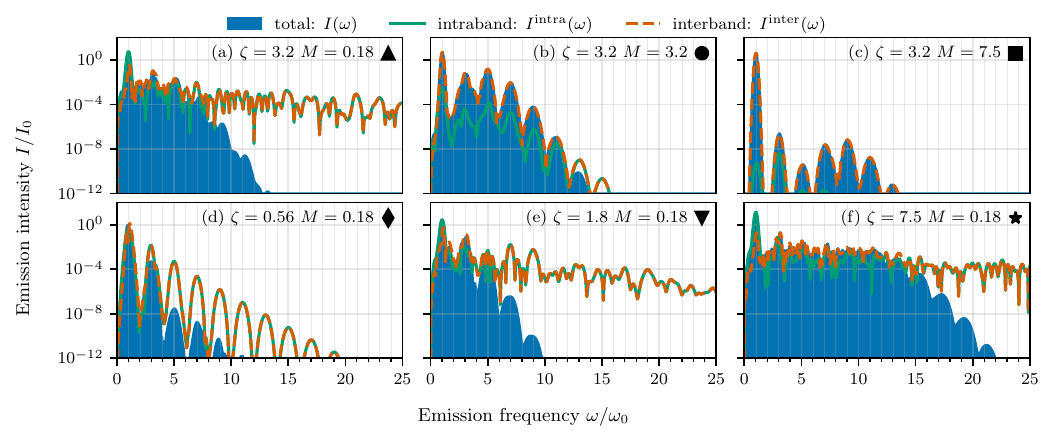}
    \caption{\label{fig:fig2} Total frequency-resolved emission intensity $I(\omega)$ (Eq.~(\ref{eq:IDecomposition}), shaded blue) compared to intra-band (solid green line) and inter-band (dashed orange line) contributions
    for different multi-photon numbers, $M$, and strong-field parameters, $\zeta$, defined in Eq.~(\ref{eq:parameters}).
    Here, we drive a massive Dirac model, \eqn(\ref{dirac hamiltonian scaled}), by the electric field in \eqn(\ref{driving field}) with $\sigma=3\pi/\omega_0$. Top row panels show intensities with different $M$ for $\zeta=3.2$, demonstrating coherent suppression (CS) due to the interference term in Eq.~(\ref{eq:IDecomposition}) (not shown) for small $M$ and inter-band dominance for large $M$. Bottom row panels depict results for various values of $\zeta$ at $M=0.18$, indicating appearance of CS for a wide range of $\zeta$. Markers refer to position in parameter space in \fig \ref{fig:fig1}.  For a driving frequency of $\omega_0/2\pi=\SI{10}{THz}$, the multi-photon numbers $M\in\{0.18,3.2,7.5\}$ correspond to bandgaps of $\Delta\in\{\SI{7.5}{meV}, \SI{130}{meV}, \SI{310}{meV}\}$. At a Fermi velocity of $v_F=\SI{5.0d5}{m/s}$, strong-field parameters $\zeta\in\{0.56,3.2,7.5\}$ correspond to peak field strengths of $E\in\{\SI{0.015}{MV/cm}, \SI{0.083}{MV/cm},\SI{0.19}{MV/cm}\}$.}
\end{figure*}

\section{Theoretical Framework}
We study a two-dimensional massive Dirac Hamiltonian
\begin{eqnarray}\label{dirac hamiltonian scaled}
    \hdirac(\bm{\kappa}) = \frac{\zeta}{2}\left(\kappa_x\hat{\sigma}_x + \kappa_y\hat{\sigma}_y \right) + \frac{M}{2}\hat{\sigma}_z ,\label{dimensionless linear Hamiltonian}
\end{eqnarray}
in dimensionless form driven by a linearly polarized electric field,
\begin{equation}\label{driving field}
    \bm{E}(t)=-\dot{\bm{A}}(t) \quad,\quad
    \bm{A}(t)=\bm{e}_x\frac{E}{\omega_0}\cos(\omega_0 t)e^{-t^2/2\sigma^2} ~,
\end{equation}
with standard deviation $\sigma$, peak field strength $E$, central angular frequency $\omega_0$, Pauli matrices $\hat{\sigma}_{x,y,z}$, and scaled wave-vector $\bm{\kappa}=\omega_0\,\bm{k}/E$.
The multi-photon number and the strong-field parameter,
\begin{equation}
    M=\Delta/\omega_0  \qand \zeta=2v_F\,E / \omega_0^2 \, ,
    \label{eq:parameters}
\end{equation}
in terms of the bandgap $\Delta$ and the Fermi velocity $v_F$ characterize the dynamics of the system \cite{Heide2021,Kruchinin2018Colloquium:Systems}.
All physical quantities above and throughout this work are given in atomic units unless stated otherwise.
The Hamiltonian above and equations of motion below were brought into dimensionless form by introducing a characteristic time scale $t_c=1/\omega_0$ and length scale $l_c=\omega_0/E$ (for details see appendix \ref{sec:A}).
We note that the massive Dirac model describes the qualitative behavior of a large class of materials.
However, it cannot capture more specific details e.g. six-fold polarization dependence in graphene \cite{Yoshikawa2017} or population asymmetries in Weyl-semimetals \cite{Bharti2023HowSemimetals}.
The evolution in dimensionless time $\tau=t/t_c=\omega_0\,t$ is governed by the well-established SBE \cite{Lindberg1988,Krieger1986,Krieger1987,Blount1962,Wilhelm2021,Schafer2002SemiconductorPhenomena},
\begin{eqnarray}\label{eq:SBE}
    &&\left[i\,\partial_{\tau} +\dfrac{i(1-\delta_{mn})}{2\pi\,\tau_2}+ \epsilon_{mn}(\bm{\kappa}_{\tau})\right]\rho_{mn}(\bm{\kappa},\tau)
    = \\
    &&\bm{F}(\tau)\cdot\sum_{r\in\{c,v\}}[\rho_{mr}(\bm{\kappa},\tau)\bm{d}_{rn}(\bm{\kappa}_{\tau}) - \bm{d}_{mr}(\bm{\kappa}_{\tau})\rho_{rn}(\bm{\kappa},\tau)]\nonumber,
\end{eqnarray}
in the adiabatic Houston basis with phenomenological dephasing time $\tau_2=T_2/t_c$,
scaled field ${\bm{F}(\tau)=\bm{E}(\tau/\omega_0)/E}$
and kinematic wavenumber
 ${\bm{\kappa}_{\tau}=\bm{\kappa}-\bm{a}(\tau)}$ with
${\bm{a} (\tau)=\frac{\omega_0}{E}\bm{A}(\tau/\omega_0)}$.
Indices $r,m$ and $n$ can take the values $c$ and $v$ for conduction and valence band states.
We adopt the initial condition ${\rho_{mn}(\tau\rightarrow -\infty) =\delta_{mn}\delta_{nv}}$ of a completely filled valence band.
The dipoles $\bm{d}_{mn}(\bm{\kappa})=i\mel{m\bm{\kappa}}{\partial_{\bm{\kappa}}}{n\bm{\kappa}}$ and energies $\varepsilon_n(\bm{\kappa})$ are defined in terms of eigenstates $\ket{n\bm{\kappa}}$ of any Hamiltonian $\hat{H}(\bm{\kappa})$,
\begin{equation}
    \hat{H}(\bm{\kappa})\ket{n\bm{\kappa}}=\varepsilon_n(\bm{\kappa})\ket{n\bm{\kappa}},
\end{equation}
and $\epsilon_{mn}(\bm{\kappa})=\varepsilon_m(\bm{\kappa})-\varepsilon_n(\bm{\kappa})$ denotes the energy differences between bands.

We note that a single calculation in our dimensionless formalism corresponds to an ensemble of physical realizations.
For example, choosing a Fermi velocity of ${v_F=\SI{5.0d5}{m/s}}$ and a driving frequency of ${\frac{\omega_0}{2\pi}=\SI{10}{THz}}$ determines all other quantities via \eqn\ref{eq:parameters}.
Using this choice, the data shown in the top panel of \fig\ref{fig:fig1} corresponds to a bandgap of ${\Delta=\SI{8.4d-4}{au}=\SI{23}{meV}}$, a peak field strength of ${E=\SI{2.9d-5}{au}=\SI{0.15}{MV/cm}}$, and a peak intensity ${I=\SI{4.4d-9}{au}=\SI{28}{MW/cm^2}}$ of the driving field.

We are interested in the frequency-resolved emission intensity calculated via Larmor's formula \cite{JohnDavidJackson1975},
\begin{equation}\label{instensity larmor}
    I(\omega) = I_0\,\omega^2\abs{\bm{j}(\omega)}^2,
\end{equation}
which is given here in terms of the natural intensity scale ${I_0=l_c^{-3}c^{-3}t_c^{-2}/3=E^3/3c^3\omega_0}$.
Furthermore, the Fourier transform $\bm{j}(\omega)$ of the dimensionless current density is given by
\begin{eqnarray}\label{eq:bzintegral}
    \bm{j}(\tau) &&= \int_{BZ}\frac{d\bm{\kappa}}{(2\pi)^2}\Tr\left[\jhat_{\bm{\kappa}}\hat{\rho}(\bm{\kappa}+\bm{a}(\tau),\tau)\right],
    \label{current density definition}
\end{eqnarray}
where the current operator $\jhat_{\bm{\kappa}}\!=\!\frac{\partial\hat{H}}{\partial\bm{\kappa}}$ is employed.
The total current can be decomposed into intra- and inter-band contributions,
\begin{eqnarray}\label{intra-band inter-band decomposition current}
    \bm{j}(\tau) &&=
    \bm{j}^{\text{intra}}(\tau) + \bm{j}^{\text{inter}}(\tau), \\
     \bm{j}^{\text{intra}}(\tau) &&=  \int_{BZ}\frac{d\bm{\kappa}}{(2\pi)^2}\sum_n \rho_{nn}(\bm{\kappa}+\bm{a}(\tau),\tau)\, j_{nn}(\bm{\kappa}),\nonumber\\
     \bm{j}^{\text{inter}}(\tau) &&=
      \int_{BZ}\frac{d\bm{\kappa}}{(2\pi)^2} \sum_{m\neq n} \rho_{mn}(\bm{\kappa}+\bm{a}(\tau),\tau)\,j_{nm}(\bm{\kappa}),\nonumber
\end{eqnarray}
with $j_{mn}(\bm{\kappa})=\mel{m\bm{\kappa}}{\jhat_{\bm{\kappa}}}{n\bm{\kappa}}$.
Correspondingly, this decomposition carries over to the spectral intensity,
\begin{eqnarray}
I(\omega)=I^{\text{intra}}&&(\omega) + I^{\text{inter}}(\omega)+ I^{\text{interference}}(\omega),
   \label{eq:IDecomposition}
\\
    I^{\text{intra/inter}}(\omega) &&= I_0\omega^2\abs{\bm{j}^{\text{intra/inter}}(\omega)}^2 \, ,
       \label{eq:IDecomposition2}
    \\
     I^{\text{interference}}(\omega) &&=
      I_0\omega^2\Re\left([\bm{j}^{\text{intra}}(\omega)]^*\bm{j}^{\text{inter}}(\omega)\right) \nonumber\\
      &&=2\sqrt{I^{\text{inter}}(\omega)I^{\text{intra}}(\omega)}\cos(\phi) \, ,
      \label{eq:IDecomposition3}
\end{eqnarray}
where the phase difference $\phi$ between inter- and intra-band currents in frequency space was defined.
We note that this choice of decomposition is not unique; for different options and discussions, see \cite{Wilhelm2021,Yue2022IntroductionTutorial,Yue2023}.
While only the \emph{total} current and intensity are physical observables, a sensible decomposition allows to interpret the underlying microscopics and generally depends on the physical regime.  We assume that the basis of the unperturbed Hamiltonian, i.e.\ the system's band structure at rest, provides the natural decomposition framework in our case.

\section{Coherent suppression of high harmonics}
To study the interplay of intra- and inter-band dynamics in the HHG signal systematically, we compute the frequency-resolved total emission, \eqn\eqref{instensity larmor}, and its decomposition, Eqs.~(\ref{eq:IDecomposition}, \ref{eq:IDecomposition2}, \ref{eq:IDecomposition3}), over an extensive parameter range\footnote{We employed in-house software to solve the SBE massively parallel on state-of-the art GPUs\cite{Utkarsh2024automated,Rackauckas2017}. Code available at \cite{code}, for newest version check \url{https://doi.org/10.5281/zenodo.17828204}.} spanned by $M$ and $\zeta$. We start by discussing results without dephasing, $\tau_2\!=\! \infty$. The total emission is highest for low frequencies and decays on the whole with increasing frequency showing the characteristic HHG peaks \cite{Ghimire2018High-harmonicSolids,Ghimire2011,Lewenstein1994, Yue2022IntroductionTutorial}, see \fig \ref{fig:fig1}, top panel, and \fig \ref{fig:fig2}. Most notably, in spectral regions where the intra- and inter-band signals contribute equally, especially at large frequencies (cf.~\fig \ref{fig:fig1}, top panel), we observe a particularly rapid decline of the total emitted intensity with frequency. We attribute this suppression effect to inter- and intra-band contributions cancelling coherently, leading to small or vanishing total emitted signal. To quantify the extent to which intra- and inter-band signals contribute equally, we compute their ratio $R=\ev{I^{\text{inter}}/I^{\text{intra}}}_{\omega}$, where $\langle .\rangle_{\omega}$ denotes the average over all frequencies with contributions $I^{\text{inter/intra}}$ above the numerical noise threshold. The lower panel of \fig  \ref{fig:fig1} demonstrates that $R$ is closest to unity, and hence enables coherent suppression (CS) of the total signal, for small $M$ or large $\zeta$.
We illustrate the different shapes of the HHG emission in different parameter regimes and their decomposition into inter- and intra-band contributions in the exemplary spectra in \fig \ref{fig:fig2}. For moderate and large $M$, the inter-band contribution dominates the total HHG emission (see \fig \ref{fig:fig2}(b,c)).

\begin{figure*}[t]
    \centering
    \includegraphics{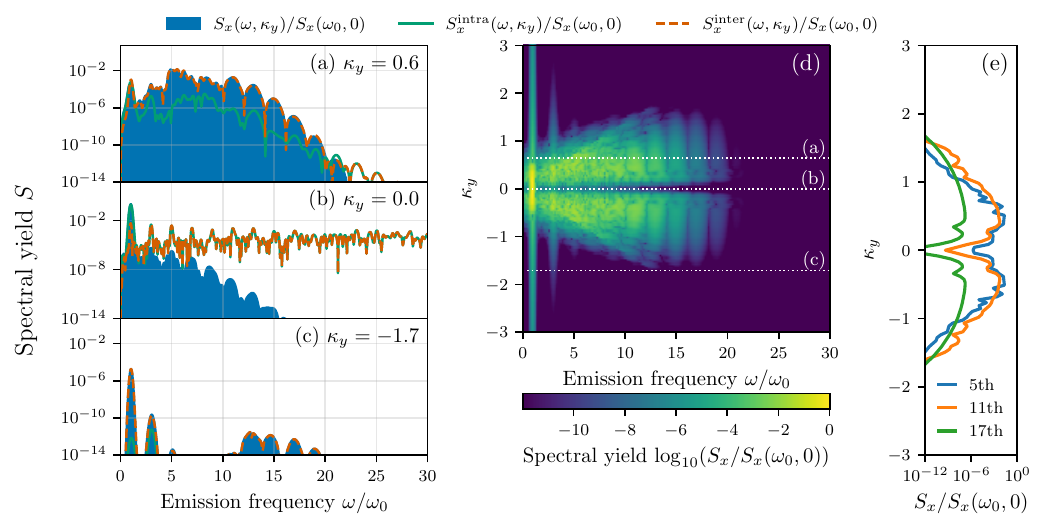}
    \caption{\label{fig:figappendix1} Spectral yields from one-dimensional slices through the Brillouin zone (see \eqn\ref{eq:kyresolved emission}) for a massive Dirac model with $\zeta = 7.5$ and $M = 0.18$ (all parameters as in \fig\ref{fig:fig2}f ). (a)-(c) show the decomposition of the total (blue shaded) emission spectrum into intra-band (solid green line) and inter-band (dashed orange line) for three different locations of the cut, $\kappa_y = 0.6$ (a), $\kappa_y = 0$ (b) and $\kappa_y = -1.7$ (c). Panel (d) displays the color-coded and frequency-resolved total emission intensity obtained from several horizontal, one-dimensional slices of the Brillouin zone integral for different $\kappa_y$. (e) shows the $\kappa_y$-dependence of the 5th, 11th, and 17th harmonic. All intensities are normalized to the first harmonic of the $\kappa_y=0$ slice for easier comparison.}
\end{figure*}

For small $M$, inter- and intra-band signals contribute equally and cancel coherently at larger $\omega$, suppressing the total HHG signal at frequencies greater than a certain threshold value, often referred to as harmonic  cutoff \cite{Vampa2014,Vampa2015,Yue2020,Yue2021,Lewenstein1994}.
Such a cutoff is absent in the inter- and intra-band contributions taken separately, each decaying smoothly down to the numerical noise level (cf. appendix \ref{sec:A2}).
This plateau depends linearly on $\zeta$ for small $M$ (see appendix \ref{sec:D}, which is in line with previous theoretical studies \cite{chizhova2017,Murakami2022} and the three-step model of HHG \footnote{A rough estimate for the maximum band gap of re-collision trajectories is $\omega_{\text{cutoff}}\approx\sqrt{\zeta^2+M^2}$.}.

However, the three-step model assumes a low depletion of the valence band and a dominant inter-band current \cite{Vampa2014,Yue2020}, which seems to contradict our finding.
A detailed investigation of the spectral emission for different parts of the Brillouin zone integral, cf.~\eqn(\ref{current density definition}), provides clarity:
consider the $\kappa_y$-resolved spectral content obtained by performing the Brillouin zone integral (\eqn(\ref{eq:bzintegral})) along $\kappa_x$ for different fixed $\kappa_y$.
This results in $\kappa_y$-dependent spectral yields,
\begin{eqnarray}\label{eq:kyresolved emission}
    S_x(\omega,\kappa_y) &&= \omega^2\abs{j_x(\omega,\kappa_y)}^2 ,\nonumber\\
    S_x^{\text{inter/intra}}(\omega,\kappa_y) &&= \omega^2\abs{j_x^{\text{inter/intra}}(\omega,\kappa_y)}^2  ,
\end{eqnarray}
of an ensemble of one-dimensional systems, which are slices in the BZ along the laser field direction.
The notation in \eqns\ref{eq:kyresolved emission} is intended to differentiate these theoretical quantities from the intensities in \eqns\ref{eq:IDecomposition}.
The spectral yields are defined in terms of current densities,
\begin{eqnarray}\label{eq:inter-intra-current-slices}
    j_x(\tau,\kappa_y) &&=
    j_x^{\text{intra}}(\tau,\kappa_y) +j_{x}^{\text{inter}}(\tau,\kappa_y), \nonumber\\
     j_x^{\text{intra}}(\tau,\kappa_y) &&=  \int_{BZ}\frac{d\kappa_x}{2\pi}\sum_n \rho_{nn}(\bm{\kappa}+\bm{a}(\tau),\tau)\, j_{x,nn}(\bm{\kappa}),\nonumber\\
     j_x^{\text{inter}}(\tau,\kappa_y) &&=
      \int_{BZ}\frac{d\kappa_x}{2\pi} \sum_{m\neq n} \rho_{mn}(\bm{\kappa}+\bm{a}(\tau),\tau)\,j_{x,nm}(\bm{\kappa}) ~,\nonumber
\end{eqnarray}
with $j_{x,mn}=\bra{m\bm{\kappa}}\pdv{\hat{H}}{\kappa_x}\ket{n\bm{\kappa}}$.
Figure \ref{fig:figappendix1}a-c show three example spectra taken at different wavenumbers $\kappa_y$  and a heatmap of intensies $I(\omega,\kappa_y)$ over the full BZ in panel d.
There, the total spectrum stemming from the 1D-line at $\kappa_y=0$ (panel b) does not exhibit a plateau-like structure, but instead features CS already beginning at the third harmonic (see \fig \ref{fig:figappendix1}b,d).
With increasing $|\kappa_y|$ the total emission spectra from these 1D  slices begin to hold a plateau which is dominated by the inter-band contribution, as it can be seen from panels (a) and (c).
This is precisely the plateau region visible in \figs\ref{fig:fig1} and \ref{fig:fig2}(f).

Physically, this can be explained in the following way: all $\bm{\kappa}$-modes are accelerated exclusively in $x$-direction, such that they encounter an effective gap of $\sqrt{M^2+\zeta^2\kappa_y^2}$ at $\kappa_x=0$.
Consequently, the one-dimensional slices experience a crossover from the small-gap/CS regime to recollision-/three step model-behavior with increasing effective gap.
This also explains, why, the low harmonics (up to third order) persist in all samples of $\kappa_y$ shown in \fig\ref{fig:figappendix1}: for increasing gaps there is a transition to the regime perturbative in the field strength.
The $\kappa_y$-dependence of the 5th, 11th and 17th harmonic are shown in \fig\ref{fig:figappendix1}e, where the dip at $\kappa_y=0$ due to CS is also visible.
Furthermore, it illustrates that the yield slightly 'spreads' with increasing harmonic, e.g. the 11th harmonic is more pronounced in slices with $|\kappa_y| > 1$.

The preceding considerations motivate us to restrict the SBE to one dimension to unravel the mechanism behind CS for small multi-photon numbers $M=\Delta/\omega_0$.
Setting $\kappa_y=0$ and performing a unitary rotation of \eqn(\ref{dirac hamiltonian scaled}) yields the effective 1D Hamiltonian,
\begin{equation}\label{eq:dirac1dhamiltonian}
\hat{H}_{\text{1d}}(\kappa_x)=\zeta\kappa_x\sigma_z/2 + M\sigma_x/2 \, .
\end{equation}
To facilitate an expansion of the SBE solution around $M=0$ we employ the diabatic basis, i.e.~the eigenstates of \eqn(\ref{eq:dirac1dhamiltonian}) for $M\!=\!0$.
Then the equations of motion remain well-defined, whereas the adiabatic Houston basis is not differentiable at $\bm{\kappa}\!=\!0$ for vanishing $M$ and thus the dipoles are ill-defined in this limit.
In physical terms, these are decoupled left- and right-movers instead of conduction/valence band charge carriers, cf.~the dashed and solid lines in the lower right panel of \fig\ref{fig:fig1}.

A change of basis of the SBE \eqref{eq:SBE} yields the equations of motion for
   $ \rho_{\pm\pm}(\kappa_x,\tau)=\mel{\pm\kappa_x}{\hat{\rho}(\tau)}{\pm\kappa_x}$.
It is sufficient to consider the dynamics of the coherence $\rho_{+-}(\kappa_x,\tau)$ and the imbalance $\delta(\kappa_x,\tau)=\frac{1}{2}(\rho_{++} -\rho_{--})$.
All matrix elements $\rho_{\pm\pm}$ then follow from $\tr\hat{\rho}=1$ and the unitarity of the density matrix.

We expand coherence and imbalance for $M\ll 1$,
\begin{eqnarray}\label{asymptotic expansion}
    \delta(\kappa_x,\tau) &\sim \frac{1}{2\varepsilon_c}\sum_{n=0} M^n \delta^{(n)}(\kappa_x,\tau), \nonumber\\
    \rho_{+-}(\kappa_x,\tau) &\sim \frac{1}{2\varepsilon_c}\sum_{n=0} M^n\rho^{(n)}_{+-}(\kappa_x,\tau),
\end{eqnarray}
with $\varepsilon_c=\varepsilon_c(\bm{\kappa})\rvert_{\bm{\kappa}=(\kappa_x,0)}$.
Solving the equations of motion  yields
for the total current density ${j_{x,1d}(\tau)=\int_{BZ}\frac{d\kappa_x}{2\pi}\Tr\qty[\pdv{\hat{H}_{1d}}{\kappa_x}\hat{\rho}(\kappa_x+a_x(\tau),\tau)]}$
the approximation
\begin{eqnarray}\label{eq:asymtotic-approximation-dirac}
    j_{x,1d}^{(0)}(\tau)
        &&=  -\int_{BZ}\frac{\dd\kappa_x}{2\pi} \frac{\zeta}{\varepsilon_c}
        \delta^{(0)}(\kappa_x+a_x(\tau),\tau) \nonumber\\
        &&= -\frac{\zeta}{2\pi}\,a_x(\tau) + \order{M} \, .
\end{eqnarray}
See  (\fig\ref{fig:figappendix2}(b)) for a comparison with corresponding numerical calculations showing quantitative agreement.
Equation (\ref{eq:asymtotic-approximation-dirac}) allows a qualitative explanation of CS:
the current $  j_{x,1d}^{(0)}(\tau)$ is a Gaussian multiplied by a cosine, see Eq.~(\ref{driving field}), yielding  a power spectrum with only one peak at $\omega_0$. Therefore high-frequency components are absent from the total emission intensity for small $M$.
Since the current operator $\jhat_{\bm{\kappa}}$ is diagonal in the diabatic basis $\ket{\pm\kappa_x}$, no off-diagonal contribution exists.
High frequencies in the individual inter- and intra-band contributions (see bottom row panels in \fig\ref{fig:fig2}), which are orders of magnitude above the total signal, result from the pronounced peaks of dipoles and current matrix elements in the complementary adiabatic basis.
Altogether we conclude that primarily a small multi-photon number $M$ is  responsible for CS in the Dirac system.

\FloatBarrier

\begin{figure}[ttt]
\includegraphics{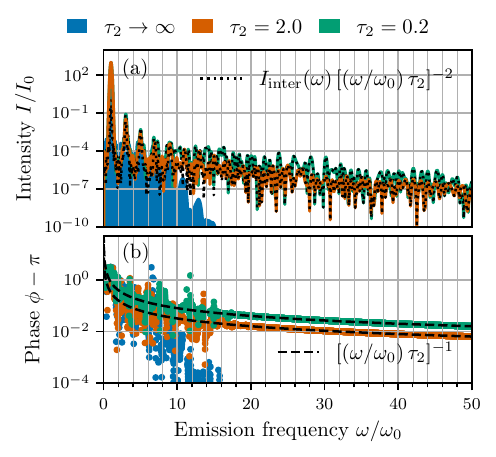}
\caption{
Dephasing-induced HHG -- (a) Emission intensity $I(\omega)$ (\eqn(\ref{eq:IDecomposition})) and (b) deviation from $\pi$ of relative phase $\phi$ between inter- and intra-band intensities (\eqn(\ref{eq:IDecomposition3})) for different dephasing strengths in the one-dimensional massive Dirac model, \eqn(\ref{eq:dirac1dhamiltonian}).
Colors correspond to scaled dephasing times ${\tau_2\rightarrow\infty}$ (blue), ${\tau_2=2.0}$ (orange) and ${\tau_2=0.2}$ (green).
The total emission (a) coincides with ${I^{\mathrm{inter}}(\omega)/[(\omega/\omega_0) \tau_2]^2}$ (dotted lines) for ${\omega\gg\omega_0}$.
This follows directly from ${\phi\!-\!\pi \approx 1/[\omega/ T_2]=1/[(\omega/\omega_0)\tau_2]}$ (dashed lines in (b)).
Parameters used are $\zeta=7.5$,$M=0.18$ and $\sigma=3\pi/\omega_0$ as in \figs~\ref{fig:fig2}(f), \ref{fig:fig4} and marked by $\blacksquare$ in \fig\ref{fig:fig1}.
For a driving frequency of $\omega_0/2\pi=\SI{10}{THz}$, the scaled dephasing times $\tau_2\in\{0.2,2\}$ correspond to $T_2\in\{\SI{20}{fs},\SI{200}{fs}\}$.}
\label{fig:fig3}
\end{figure}

\section{Dephasing and re-entrant HHG}
We now include dephasing by means of the dephasing time $\tau_2=\omega_0 T_2 / 2\pi$ in the SBE (\ref{eq:SBE}).
To study its effect on CS we first consider the relevant one-dimensional model introduced above.
Figure~\ref{fig:fig3}(a) shows the emission intensity for different $\tau_2$ in the parameter regime of CS.

We observe a \emph{re-entrance} of high harmonics at $\omega \gg \omega_0$ for finite $\tau_2$.
This counterintuitive behavior can be traced back to the $\tau_2$-dependence of the relative phase $\phi$ of the spectral inter- and intra-band currents (see \eqn(\ref{eq:IDecomposition3})).
As shown in \fig\ref{fig:fig3}(b), for dephasing times far beyond the laser cycle ($\tau_2\! \gg \! 1$)  the relative phase is $\phi(\omega) \approx \pi$ as expected for destructive interference and leading to CS.
However, for a dephasing comparable to the laser cycle, we find $\phi(\omega)\!-\!\pi \approx 1/\omega T_2= 1/[(\omega/\omega_0)\tau_2]$ (cf.~\fig\ref{fig:fig3}(b)).
As a result, the destructive interference is disturbed and a total HHG signal emerges for higher frequencies.

Intra- and inter-band contributions obey $I^{\mathrm{inter}}\approx I^{\mathrm{intra}}$ for $\omega \gg\omega_0$ both for short and long dephasing times.
Combining this fact with $\cos\phi(\omega\!\gg\!\omega_0)\approx \!-1\!+\!1/2(\omega T_2)^2$ in
\eqn(\ref{eq:IDecomposition3})
reveals that total emission intensity
follows $(\omega T_2)^{-2}\,I^{\mathrm{inter}}(\omega)$. This is demonstrated numerically in \fig\ref{fig:fig3}(a).

\begin{figure}[t]
\includegraphics{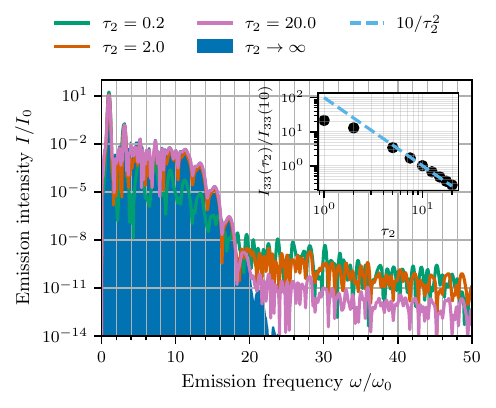}
\caption{Effect of dephasing on high harmonic emission intensity $I(\omega)$ for the driven two-dimensional massive Dirac model for dephasing times $\tau_2\rightarrow\infty$ (blue), $\tau_2=20.0$ (purple),$\tau_2=2.0$ (orange), and $\tau_2=0.2$ (green) in units of the laser cycle $2\pi/\omega_0$.
Re-entrance of high harmonics beyond the plateau is clearly visible, but less pronounced compared to the one-dimensional model (see \fig\ref{fig:fig3}).
Parameters are $\zeta=7.5$, $M=0.18$, and $\sigma=3\pi/\omega_0$ as in \figs\ref{fig:fig2}f), \ref{fig:fig3} and marked by $\blacksquare$ in \fig\ref{fig:fig1}).
The inset shows the dependence of the 33rd harmonic as a function of dephasing time $\tau_2$, normalized by the 33rd harmonic for $\tau_2=10$.
For a driving frequency of $\omega_0/2\pi=\SI{10}{THz}$, the scaled dephasing times $\tau_2\in\{0.2,2,20\}$ correspond to $T_2\in\{\SI{20}{fs},\SI{200}{fs},\SI{2}{ps}\}$.}
\label{fig:fig4}
\end{figure}

In the following, we examine whether the re-entrance mechanism governs the full 2D dynamics.
Figure \ref{fig:fig4} shows the total intensity of HHG emission  for different values of the dephasing time, indicating two opposite trends:

the pronounced HHG plateau originating from the $\bm{\kappa}$-modes with $\kappa_y\neq 0$ gradually disappears with increasing dephasing strength.
This behavior is comparable to \cite{Heide2022ProbingGeneration2}, where inter-band-dominated harmonics weaken with increasing dephasing\footnote{We note that in \cite{Heide2022ProbingGeneration2} a larger multi-photon number ($M\approx 7$) was investigated. Additionally, the massive Dirac Hamiltonian is only a rough qualitative model for the experiment in \cite{Heide2022ProbingGeneration2}.}.
At the same time, higher harmonics emerge beyond the plateau in \fig\ref{fig:fig4}, which are created via the re-entrance mechanism described above.
They are particularly pronounced and survive even for strong dephasing.
The inset in \fig\ref{fig:fig4} shows the variation of the 33rd harmonic with dephasing strength.
It clearly follows the fit $\propto1/\tau_2^2$ for weak dephasing with $\tau_2 \gtrsim 2.0$, but for shorter $\tau_2$ we observe deviations.
Therefore, the influence of dephasing goes beyond affecting solely the relative phase $\phi(\omega)$ of inter- and intra-band emission.
Using techniques like resonant photo-doping, it is possible to vary the dephasing strength \cite{Heide2022ProbingGeneration2}, which could allow a distinction between inter-band dominated and re-entrant harmonics or an extraction of dephasing time.

Note that $\tau_2=0.2$ corresponds to $20\,$fs at a driving frequency of $10\,$THz, comparable to simulations, e.g., in \cite{Schmid2021,Hohenleutner2015,Vampa2014,Yue2021,Floss2018AbSolids,Yue2021ExpandedRecollisions}.
The model used to describe dephasing in the SBE (\ref{eq:SBE})
is applicable to a wide range of systems: it can mimic propagation-induced decoherence in the bulk \cite{Floss2018AbSolids,Kilen2020} as well as various many-body effects such as electron-electron or polarization-polarization scattering~\cite{Hohenleutner2015,Wilhelm2021}.

\section{Coherent suppression for Bilayer graphene model}

\begin{figure*}[t]
    \centering
    \includegraphics{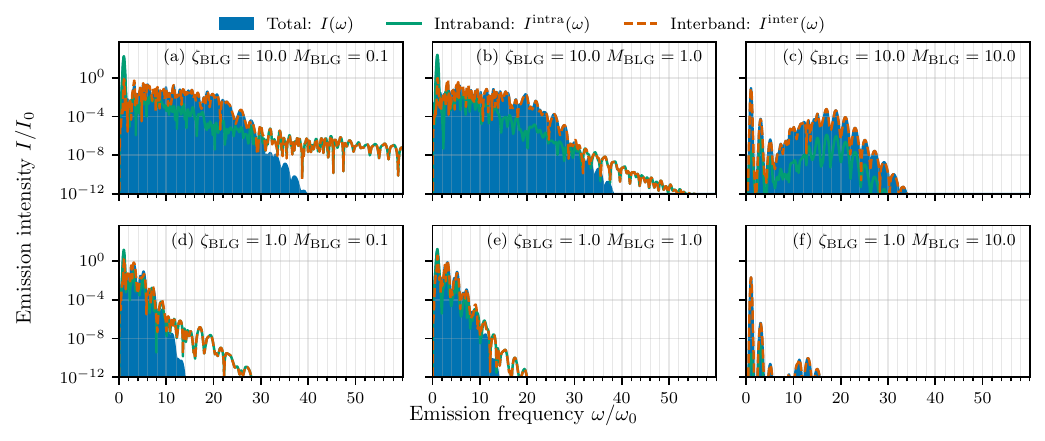}
    \caption{\label{fig:figappendix3} HHG for a bilayer graphene (BLG) model. Total emission intensity $I(\omega)$ (Eq.~(\ref{eq:IDecomposition}), shaded blue) compared to intra-band (solid green line) and inter-band (dashed orange line) contributions
    for different multi-photon numbers, $M_{\text{BLG}}$, and strong-field parameters, $\zeta_{\text{BLG}}$, defined in Eq.~(\ref{eq:parameters}).
    Here, we drive a toy model of bilayer graphene, \eqn(\ref{blg hamiltonian scaled}), by the electric field in \eqn(\ref{driving field}) with $\sigma=3\pi/\omega_0$. We show different $M_{\text{BLG}}$ for $\zeta_{\text{BLG}}=10$ in the top row panels and for $\zeta_{\text{BLG}}=1$ in the bottom row panels. The emission intensity behaves qualitatively similar to the Dirac model (cf. \fig\ref{fig:fig2}): panels (a), (b), (d) and (e) with  $M_{\text{BLG}}\in\{0.1,1\}$ exhibit coherent suppression which is less pronounced.
    For a driving frequency of $\omega_0/2\pi=\SI{10}{THz}$, the multi-photon numbers $M_{\mathrm{BLG}}\in\{0.1,1,10\}$ correspond band gaps of $\Delta_{\mathrm{BLG}}\in\{\SI{4.2}{meV},\SI{42}{meV},\SI{420}{meV}\}$.
    Additionally setting the effective mass as $m=0.032m_e$, the strong-field parameters $\zeta\in\{1,10\}$ correspond to field strengths of $E\in\{\SI{0.054}{MV/cm},\SI{0.17}{MV/cm}\}$.}
\end{figure*}

It was hypothesized \cite{Wilhelm2021} that a linear dispersion is key for CS.
To investigate this we study a toy model of bilayer graphene (BLG),
\begin{align}\label{blg hamiltonian scaled}
   \hblg(\bm{\kappa}) = \frac{\zeta_{\text{BLG}}}{2}[(\kappa_y^2-\kappa_x^2)\hat{\sigma}_x - 2\,\kappa_x \kappa_y \hat{\sigma}_y] + \frac{M_{\text{BLG}}}{2}\hat{\sigma}_z ,
\end{align}
driven by the pulse defined in \eqn(\ref{driving field}).
The multi-photon number and strong-field parameter,
\begin{align}
    M_{\text{BLG}}=\Delta_{\text{BLG}}/\omega_0 \qc
    \zeta_{\text{BLG}}=E^2/m\,\omega_0^3 ,
\end{align}
are expressed in terms of bandgap $\Delta_{\text{BLG}}$ and effective mass $m$.
This model describes massive chiral electrons with an added momentum-independent gap \cite{McCann2013}.
Similarly to the massive Dirac Hamiltonian, CS is present for small multi-photon numbers, even if not as pronounced (see \fig\ref{fig:figappendix3}): for high frequencies inter- and intra-band contributions, $I^{\mathrm{inter}}(\omega)$ and $I^{\mathrm{intra}}(\omega)$, are orders of magnitude larger than the total emission intensity $I(\omega)$ (numerically zero).
In comparison to the Dirac model this difference is smaller and inter- and intra-band contributions decay faster.
The qualitative explanation for CS is analogous to the Dirac model: $\bm{\kappa}$-modes around $\kappa_y=0$ are responsible for large intra- and inter-band current of same magnitude and opposite phase for large frequencies.
These contributions origin from rapidly changing current matrix elements around $\bm{\kappa}=0$, which are smoother in case of the BLG model as compared to the massive Dirac system.
Thus the interference effect is less pronounced overall.

\section{Conclusions}
We have demonstrated that the high-harmonic emission signal of driven massive Dirac fermions is strongly suppressed due to destructive interference of intra- and inter-band contributions.
We could separate the modes in the BZ responsible for the HHG plateau from those close to the band gap causing CS.
Based on an effective one-dimensional model the underlying suppression mechanism can be straightforwardly understood analytically by invoking diabatic left- and right-movers to describe the dynamics.

This coherent suppression effect and its sensitivity to dephasing primarily requires a small gap i.e. a small multi-photon number $M=\Delta/\omega_0$. Generalizing previous numerical observations \cite{Murakami2022,Wilhelm2021} we then expect CS to be apparent in a large number of materials that can be described by a weakly gapped massive Dirac model, such as graphene and topological insulator surface states \cite{Baykusheva2021,Schmid2021,Liu2010ModelInsulators,Murakami2022}.
Moreover, our extensive numerical simulations  for a BLG model provide further evidence that the mechanism of CS is qualitatively independent of the band shape.  The latter may however have quantitative consequences on the high harmonic signals.

While inter-band processes are often considered dominant in HHG from solids and 2D materials \cite{Vampa2014,Vampa2015,Yue2020,Yue2021,Zurron2018TheoryGraphene,Parks2020WannierSemiconductors}, several studies highlight the relevance of both intra- and inter-band contributions \cite{Schubert2014,Hohenleutner2015,Schmid2021,Murakami2022,Wilhelm2021}. Our results show that CS —by its nature— rules out inter-band dominance (for small gaps) and indicates that the multi-photon number $M$ plays a key role in setting the relative weight of intra- and inter-band contributions.

Counterintuitively, the HHG signal is recovered in the presence of dephasing.
This can be traced back to a fundamental power-law behavior of the relative phase between intra- and inter-band currents in frequency space.
We demonstrated that these re-entrant harmonics behave opposite to those in the inter-band dominated plateau, following a characteristic $1/\omega T_2$-dependence, which may be accessed via photodoping experiments.

\section{Data availability}
The data that support the findings of this article are openly available \cite{figdata,code} 

\section{Acknowledgments}
We thank P. Hommelhoff, A. Seith and J. Wilhelm for valuable discussions and V. Junk for useful conversations at an early stage of the project.
The work was funded by the
Deutsche Forschungsgemeinschaft (DFG, German Research Foundation) within Project-ID 314695032 – SFB 1277 and Project-ID 502572516 - GRK 2905. We acknowledge further support from the Regensburg Center for Ultrafast Nanoscopy (RUN).
The authors gratefully acknowledge the scientific support and HPC resources provided by the Erlangen National High Performance Computing Center (NHR@FAU) of the Friedrich-Alexander-Universität Erlangen-Nürnberg (FAU) under the NHR project b228da. NHR funding is provided by federal and Bavarian state authorities. NHR@FAU hardware is partially funded by the German Research Foundation (DFG) – 440719683.

\appendix

\section{Dimensionless Hamiltoninan and SBE}\label{sec:A}
In atomic units, the SBE take the form
\begin{eqnarray}\label{eq:SBEau}
    &&\left[i\,\partial_t +\dfrac{i(1-\delta_{mn})}{T_2}+  E_{mn}(\bm{k}_t)\right]\varrho_{mn}(\bm{k},t)
    = \\ \nonumber
    &&\bm{E}(t)\cdot\sum_r[\varrho_{mr}(\bm{k},t)\bm{D}_{rn}(\bm{k}_t) - \bm{D}_{mr}(\bm{k}_t)\varrho_{rn}(\bm{k},t)] ~,
\end{eqnarray}
where $\bm{E}(t)$ is the electric field, $\bm{k}_t=\bm{k}-\bm{A}(t)$ the kinematic wavenumber in terms of the vector potential $\bm{A}(t)$, and $T_2$ the phenomenological dephasing time.

The indices $m, r, n$ label the system’s bands, which remain unspecified as the rescaling applies to an arbitrary number of bands. The dipoles, \[{\bm{D}_{mn}(\bm{k})=i\mel{m\bm{k}}{\partial_{\bm{k}}}{n\bm{k}} ~,}\] and density matrix elements, \[{\varrho_{mn}(\bm{k},t)=\mel{m\bm{k}}{\hat{\rho}(t)}{n\bm{k}}} ~,\] are defined via the Bloch eigenstates $\ket{n\bm{k}}$, which solve  ${\hat{\mathcal{H}}_B(\bm{k})\ket{n\bm{k}} = E_n(\bm{k})\ket{n\bm{k}}}$.
Additionally, ${E_{mn}(\bm{k})=E_m(\bm{k})-E_n(\bm{k})}$ denotes the band energy difference.
Here $\hat{\mathcal{H}}_B(\bm{k})$ refers to any Bloch-type Hamiltonian form in atomic units.

To transform \eqn\ref{eq:SBEau} and the Hamiltonian, we introduce characteristic time and length scales, $t_c$ and $l_c$, with the scaled wavevector $\bm{\kappa} = \bm{k}l_c$.
Applying $\partial_t = \frac{1}{t_c} \partial_{\tau}$, \eqn\ref{eq:SBEau} retains its form in the scaled variables,
\begin{eqnarray}\label{eq:SBEcaled}
    &&\left[i\,\partial_{\tau} +\dfrac{i(1-\delta_{mn})}{\tau_2}+  \epsilon_{mn}(\bm{\kappa}_{\tau})\right]\rho_{mn}(\bm{\kappa},\tau)
    = \\ \nonumber
    &&\bm{F}(\tau)\cdot\sum_r[\rho_{mr}(\bm{\kappa},\tau)\bm{d}_{rn}(\bm{\kappa}_{\tau}) - \bm{d}_{mr}(\bm{\kappa}_{\tau})\rho_{rn}(\bm{\kappa},\tau)] ~,
\end{eqnarray}
where we introduced the scaled quantities,
\begin{eqnarray}\label{eq:scaledquantities}
    \rho_{mn}(\bm{\kappa},\tau) &&=\varrho_{mn}(\bm{\kappa}/l_c,\tau\,t_c) \qc
    \tau_2 = T_2/t_c, \nonumber\\
    \epsilon_{mn}(\bm{\kappa}) &&= E_{mn}(\bm{\kappa}/l_c)/t_c \qc
    \bm{\kappa}_{\tau} = \bm{\kappa}-\bm{a}(\tau) ,\nonumber\\
    \bm{F}(\tau) &&= t_c\,l_c\bm{E}(\tau\,t_c) \qc
    \bm{a}(\tau) = l_c\bm{A}(\tau\,t_c).
\end{eqnarray}
Any Hamiltonian $\hat{\mathcal{H}(\bm{k})}$ transforms according to
\begin{equation}\label{eq:scaledhamiltonoperator}
    \hat{H}(\bm{\kappa})=\hat{\mathcal{H}}(\bm{\kappa}/l_c)/t_c.
\end{equation}
The \eqns\ref{eq:scaledquantities} and \ref{eq:scaledhamiltonoperator} are valid for general $t_c$ and $l_c$, but throughout this work we use $t_c = 1/\omega_0$ and $l_c = \omega_0/E$
Hence, the massive Dirac model,
\begin{eqnarray}\label{massive Dirac au}
    \hat{\mathcal{H}}(\bm{k})=v_F(k_x\sigma_x+k_y\sigma_y)+m\sigma_z,
\end{eqnarray}
transforms to \eqn\ref{dirac hamiltonian scaled} from the main text.

\section{High-frequency behavior of massive Dirac HHG}\label{sec:A2}

\begin{figure}[h]
    \centering
    \includegraphics[width=246pt]{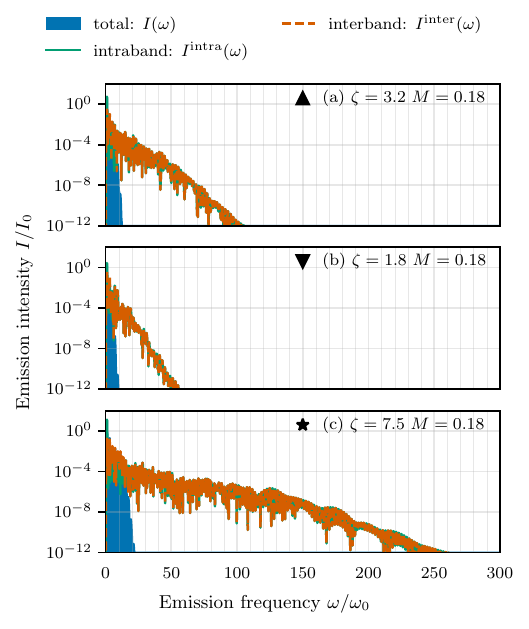} \caption{\label{fig:fig2_ex}
    High-frequency HHG of massive Dirac model: frequency-resolved emission intensity $I(\omega)$ (Eq.~(\ref{eq:IDecomposition}), shaded blue), along with its intra-band (solid green line) and inter-band (dashed orange line) components, shown for various multi-photon numbers $M$ and strong-field parameters $\zeta$ defined in Eq.~(\ref{eq:parameters}). The system is a massive Dirac model [Eq.~(\ref{dirac hamiltonian scaled})] driven by the electric field of Eq.~(\ref{driving field}) with $\sigma = 3\pi/\omega_0$. The same data as in \fig\ref{fig:fig2}a,e, and f are shown for larger frequencies. Markers indicate the corresponding points in parameter space shown in \fig~\ref{fig:fig1}.}
    \label{fig:fig2extended}
\end{figure}

For completeness, we show higher frequencies of the data from \fig\ref{fig:fig2}a,e, and f in \fig\ref{fig:fig2extended}.
This shows the lack of any sharp cutoff in the inter- and intra-band contributions, which decay smoothly down to the numerical noise threshold.
The extended frequency range in \fig\ref{fig:fig2extended} also reveals that larger strong-field parameters $\zeta$ lead to slower decay of inter- and intra-band emission, which is not clearly visible in \fig\ref{fig:fig2}.

\section{Details of asymptotic expansion}\label{sec:C}

The SBE for the massive Dirac model in the diabatic basis, i.e., the eigenstates of \(\hat{H}_0 = \zeta\kappa_x\sigma_x/2\), for \(\tau_2 \to \infty\) and \(\kappa_y = 0\) are given by
\begin{eqnarray}\label{eq:SBE-diabatic1d}
    \dot{\rho}_{+-}(\tau) &&= -2i\zeta[\kappa_x-a_x(\tau)]\rho_{+-}(\tau) + 2iM\,\delta(\tau),\nonumber\\
    \dot{\delta}(\tau) &&= -M\,\Im\rho_{+-}(\tau),
\end{eqnarray}
with initial conditions
\begin{eqnarray}
    \rho_{+-}(t\to -\infty) &&= -M/2\varepsilon_c(\kappa_x\bm{e}_x) ,\\
    \delta(t\to -\infty) &&= -\zeta \kappa_x/2\varepsilon_c(\kappa_x\bm{e}_x) .
\end{eqnarray}
For clarity, the explicit momentum dependence, \(\rho_{+-}(\kappa_x,\tau) \equiv \rho_{+-}(\tau)\) and \(\delta(\kappa_x,\tau) \equiv \delta(\tau)\), is suppressed. The expansions in \eqns 13 of the main text share the same denominator as these initial conditions, incorporating \(M\) non-perturbatively. This ensures exact matching conditions,
\begin{eqnarray}\label{eq:asymptotic-ics}
    \delta^{(n)}(\tau\rightarrow -\infty) &&= \begin{cases}
        -\zeta\kappa_x & \text{if } n=0 \\
        0 & \text{otherwise}
    \end{cases}\\
    \rho_{+-}^{(n)}(\tau\rightarrow -\infty) &&= \begin{cases}
        -1 & \text{if }n=1\\
        0 & \text{otherwise,}
    \end{cases}
\end{eqnarray}
and preserves a well-ordered expansion in \(M\) for each $\kappa_x$.
The solutions to \eqns\ref{eq:SBE-diabatic1d} and \ref{eq:asymptotic-ics},
\begin{eqnarray}
    \delta^{(0)}(\tau) &&= -\zeta \kappa_x, \quad
    \delta^{(1)}(\tau) = \rho_{+-}^{(0)}(\tau) = 0,\nonumber\\
    \rho_{+-}^{(1)}(\tau) &&=   -e^{-i\phi(\tau)}
        \left[ 1+2i\zeta\kappa_x\int^\tau_{-\infty}\dd\tau' e^{i\phi(\tau')}\right], \\
        \delta^{(n+1)}(t) &&= -\int_{-\infty}^\tau\dd\tau' \Im\rho_{+-}^{(n)}(\tau') ,\\
        \rho_{+-}^{(n+1)} &&= 2i e^{-i\phi(\tau)}\int_{-\infty}^\tau\dd\tau'\delta^{(n)}(\tau')e^{i\phi(\tau')} ~,
\end{eqnarray}
for $n\geq 1$ can be expressed in terms of the diabatic phase,
\begin{equation}
    \phi(\tau) = \zeta\int^{\tau} [\kappa_x-a_x(\tau')]\dd\tau'.
\end{equation}
Formally, matching at $\tau\rightarrow -\infty$ makes the divergent.
However, in practice this is irrelevant, because one simply matches at some time $\tau\rightarrow\tau_0\ll 0$ for which the driving field is negligible e.g.~$\tau_0=5\sigma$.
Physically, $\phi(\tau)$ the phase accumulated by a state traveling along a diabatic branch (dashed lines in \fig 1 of the main text), i.e. the phase of right- and left-movers.

The expansion above correctly reproduces the trivial exact solution for \(M = 0\) in the one-dimensional massive Dirac system.
Furthermore it reduces to conventional perturbation theory for $\kappa_x=0$, whereas for $\kappa_x\neq 0$ the non-perturbative nature of the prefactor is important, which is illustrated by a comparison in the end matter.
Using the current operator $\jhat_{\bm{\kappa}}=\zeta\sigma_z/2$ in the diabatic basis we recover the approximation given in \eqn (14) of the main text.
This approximation is valid up to $\order{M}$, because $\rho_{+-}$ does not contribute to $j_x$ and $\delta^{(1)}$,  vanishes, leaving only $\delta^{(0)}$.

\begin{figure}[thb]
    \centering
    \includegraphics[width=246pt]{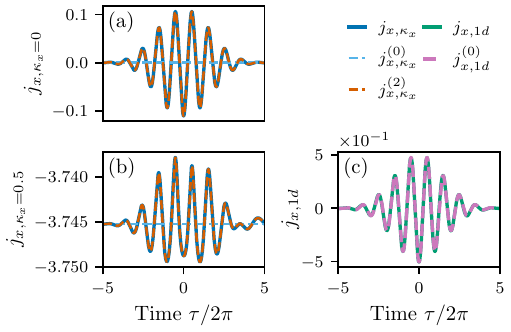}
    \caption{Comparison of asymptotic expansion (dashed lines) from \eqn (\ref{asymptotic expansion}) to numerics (solid lines) for the massive Dirac model with $\zeta=7.5$, $M=0.18$ and $\sigma=3\pi/\omega_0$. Time-resolved currents $j_{x,\kappa_x}$ for the mode $\kappa_x=0$ and $\kappa_x=0.5$ shown in panels (a) and (b), respectively, and current density $j_{x,1d}$ for the 1D system in (c).
    Superscripts indicate the order in $M$ consistent with \eqns (\ref{asymptotic expansion}) and (\ref{eq:asymptotic-approximation-singlemode-dirac}). First order is omitted, because it coincides with the leading order.}
    \label{fig:figappendix2}
\end{figure}

Here we numerically demonstrate the validity of the asymptotic expansion, \eqn(\ref{asymptotic expansion}) in \fig\ref{fig:figappendix2}.
There, we show a comparison of single-mode currents,
\begin{eqnarray}\label{eq:asymptotic-approximation-singlemode-dirac}
    j_{x,\kappa_x}^{(n)}(\tau) &&= -\frac{\zeta}{\varepsilon_c}\sum_{m=0}^n M^m \delta^{(m)}(\kappa_x,\tau) ,
\end{eqnarray}
and their integral over the BZ (see \eqn (\ref{eq:asymtotic-approximation-dirac})) for different orders in $M$.

Physically, due to the small gap $M$, valence electrons tunnel with probability close to one, hence are approximated well by decoupled left- and right movers (diabats, see \fig \ref{fig:fig1} bottom right).
The single mode at the Dirac point (\fig \ref{fig:figappendix2}a) requires second order to be sensibly approximated, whereas for $\kappa_x=0.5$ (\fig \ref{fig:figappendix2}b) the leading order is sufficient. The reason is that the latter mode is further away from the gap and thus is only weakly affected by it.
Upon integration, the oscillations captured by the second order play little role, which is demonstrated in \fig \ref{fig:figappendix2}b: the current density is approximated very well already at leading order, only the HHG spectrum reveals small higher order contributions (see \fig\ref{fig:fig3}b).

\section{Large Strong-field parameters}\label{sec:D}
\begin{figure*}[th]
    \centering
    \includegraphics{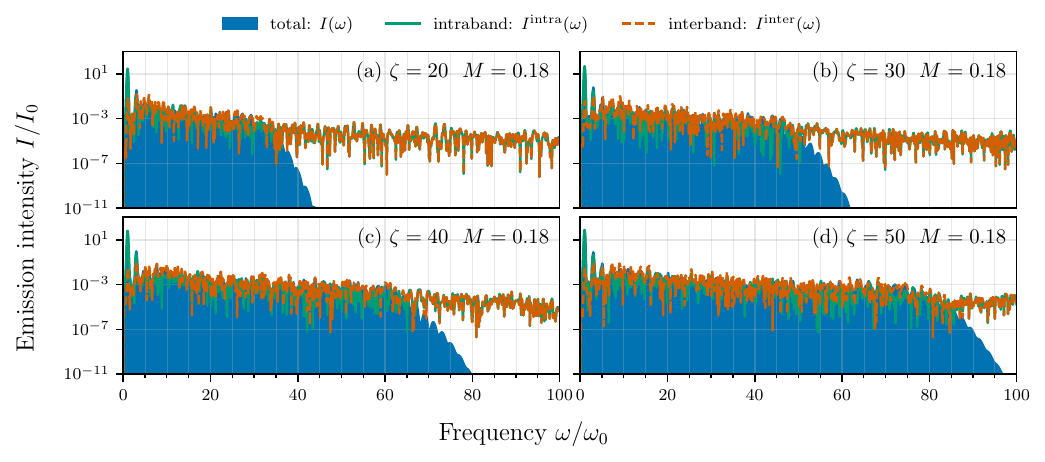}
    \caption{\label{fig:figsup3}Decomposition of frequency-resolved emission intensity $I(\omega)$ (blue shaded) into intra-band (solid green line) and inter-band (dashed orange line) according  to \eqn 8 in the main text for different values of the strong-field parameter $\zeta$. Multiphoton number $M=0.18$ and driving field of \eqn 2 in the main text with $\sigma=3\pi/\omega_0$ is used everywhere. (a) $\zeta=20 $. (b) $\zeta=30$. (c) $\zeta=40$. (d) $\zeta=50$.
    For a driving frequency of $\omega_0/2\pi=\SI{10}{THz}$, the multi-photon numbers $M=0.18$ corresponds to bandgaps of $\Delta=\SI{7.5}{meV}$.
    Additionally setting $v_F=\SI{5.0d5}{m/s}$, strong-field parameters $\zeta\in\{20,30,40\}$ correspond to peak field strengths of $E\in\{\SI{0.52}{MV/cm}, \SI{0.78}{MV/cm},\SI{1.0}{MV/cm}\}$.}
\end{figure*}
With experiments having access to field-strengths of several to several tens of $MV/cm$ \cite{Schmid2021,Heide2022ProbingGeneration}, strong-field parameters of well beyond $\zeta=10$ are possible.
This motivates an investigation of coherent suppression in that regime.
\fig\ref{fig:figsup3} shows HHG emission spectra for the massive Dirac model with $\zeta\in[20,30,40,50]$, $M=0.18$.
Coherent suppression is present in all four examples, however, we observe a shifting plateau of the total emission which is linear in $\zeta$.

\newpage

\end{document}